\documentclass[12pt]{article}
\usepackage{psfrag}
\usepackage{graphicx}
\usepackage{epstopdf}
\usepackage{amsmath}
\usepackage[T1]{fontenc}
\usepackage[utf8]{inputenc}

\begin{document}

\title{Bose-Einstein condensation in heavy ion collisions: importance of and uncertainties in the finite volume corrections}
\date{\today}
\author{Kacper Zalewski\\The H. Niewodniczański Institute of Nuclear Physics, Polish Academy of Sciences \\ Radzikowskiego 152, 31-342 Kraków, Poland \\ and \\The M. Smoluchowski Institute of Physics, Jagiellonian University \\ Łojasiewicza 11, 30-348 Kraków, Poland}
\maketitle
\begin{abstract}
The density of the Bose-Einstein condensate for non interacting pions in a cubic box, at given temperature and (average) total pion density, is calculated for three sets of  boundary conditions. The densities obtained are much larger than predicted from thermodynamics and depend significantly on the choice of the boundary conditions, even for volumes as large as $6.4\cdot  10^4$fm$^3$.
\end{abstract}

\section{Introduction}

It is plausible that a fraction of the pions produced in high-energy heavy ion collisions forms a Bose-Einstein condensate. At LHC energies, the presence of the pion condensate could, perhaps, explain the observed surplus of pions, in particular of pions with low transverse momenta, as compared to e.g. protons. According to the analysis of Begun and Florkowski \cite{BEF}, at $\sqrt{s_{NN}}= 2.76$TeV about five percent of the pions are in the condensate.

Assuming that the pions are just one kind of point-like, noninteracting bosons and that the thermodynamic limit can be used, one gets the standard formula for the density of the pions not in the condensate:

\begin{equation}\label{denter}
  \rho_{th}^*(T,\mu) = \frac{1}{2\pi^2}\int \;\frac{p^2dp}{e^{\beta(E(p)-\mu)}-1}.
\end{equation}
When the chemical potential $\mu$ exceeds the lowest single particle energy level, $E_0$, the particle density (the integrand) becomes negative at $E(\textbf{p}) = E_0$. Therefore, the chemical potential must satisfy the condition:

\begin{equation}\label{}
  \mu \leq E_0.
\end{equation}
The density (\ref{denter}) is an increasing function of the chemical potential and reaches its maximum allowed value, at given temperature, for $\mu=E_0$. If the total density of pions exceeds this maximum, the surplus pions are assumed to form the condensate. This can be justified by the formula for the density of pions in the ground state:

\begin{equation}\label{conden}
  \rho_c(T,\mu,L) = \frac{1}{L^3}\frac{g_0}{e^{\beta(E_0-\mu)}-1},
\end{equation}
where $L^3$ is the volume of the pion fluid and $g_0$ is the degeneracy of the ground state. Indeed, for $\mu \rightarrow E_0$ the ground state can accommodate an arbitrary number of pions. Let us note, however, that this justification goes beyond standard thermodynamics, where neither the density of the condensate nor the chemical potential depend on the volume. In order to get a finite, non-zero density of the condensate from (\ref{conden}), it is necessary to assume, for example, that in the vicinity of the critical temperature:

\begin{equation}\label{}
  \mu = E_0 - \frac{c(T)}{L^3}; \qquad c(T) > 0.
\end{equation}
For large volumes, the volume dependent correction to the chemical potential is negligible and the  density of particles not in the condensate can be safely calculated from (\ref{denter}) with $\mu = E_0$. This justifies, in the limit $L \rightarrow \infty$, the thermodynamic result. However, in heavy ion collisions the volume of the pion fluid is not very large. Therefore, the question arises: what are the finite volume corrections?

 The finite volume corrections to the condensate density have been estimated by Begun and Gorenstein \cite{BEG} for the free pion fluid, and, using essentially the same method, for the pion fluid in a uniform magnetic filed by Ayala, Mercado and Villavicencio \cite{AMV}.

 In the present paper we consider non-interacting pions, of a single kind, and solve the problem for three models. In each, the fluid of non interacting  pions is contained in a cubic box of volume $L^3$. In the first model, the boundary conditions are periodic. This is the approach  considered in \cite{BEG}. In the second, the particle wave functions of the pions are assumed to vanish at the boundaries of the cube. This model seems just as plausible as the previous one. In the third, the boundary conditions are free, i.e. the energy spectrum is as for free particles and formula (\ref{denter}) remains valid. This model may seem unorthodox. However, as is well known, the pion fluid rapidly expands. For expanding systems a small volume does not necessarily imply a discrete energy spectrum. A well-known example are the free Gaussian wave packets. This is the approach used in \cite{BEG2} and \cite{BEF}.

 We find in all the cases that the thermodynamic approximation grossly underestimates the amount of the condensate at given temperature and overall pion density.  The calculations show that significant differences between the exact results and the thermodynamic approximation persist up to the highest values of $L$ considered here, i.e. up to $L=40$fm. The differences between the predictions of the models depend on $L$, but in general they are very large. Therefore, if the system of non-interacting pions in a box is a good guide to estimate the condensate density, the finite volume corrections are very important and the choice of the boundary conditions is crucial for their evaluation.

\section{The models}

A full calculation, which would have to include the entire pion isotriplet and the $\pi-\pi$ interactions, is beyond the scope of the present paper. What we further call pions are spinless particles of an ideal Bose fluid with particle mass $m= 140$MeV. We also assume $g_0=1$, which is true for all the models considered here.

Let us consider the fluid of pions in a cubic box of volume $L^3$. As independent parameters we choose the edge length $L$, the temperature $T$ and the chemical potential $\mu$. The results are given for the interval

\begin{equation}\label{}
  0 \leq L \leq 40\mbox{fm}.
\end{equation}
For large $L$ it would be more realistic to consider a parallepiped, since the transverse cross-section of the fluid is certainly much less than $1600$fm$^2$, but for our qualitative discussion the cube is good enough. Moreover, reducing the transverse dimensions at given length $L$ can only increase the deviations from the thermodynamic limit.

What temperature is considered realistic depends on the theoretical approach. We have done the calculations for

\begin{equation}\label{}
  T = 100\mbox{MeV}
\end{equation}
which, according to the Alice collaboration \cite{ALI}, is reasonable for about 40\% of the most central collisions at LHC energies. We have also checked that for $T=150$MeV, which is in the range considered in \cite{BEF}, similar results are obtained, though the deviations from the thermodynamic limit are somewhat smaller.

The chemical potential is related to the total particle density. It is convenient to present the results at an $L$-independent total density of the pion fluid. The exact value of this density is not very important, but it should not be too unrealistic. We choose this density as follows. Putting $\mu = E_0$ we can calculate from (\ref{denter}) the density of the non-condensate particles in the large volume limit, i.e. in the thermodynamic approximation. Let us denote it $\rho_{th}^*(T)$. By analogy with  \cite{BEF} we assume that, in the thermodynamic approximation, this is 95\% of the total density. Thus, the total pion density is

\begin{equation}\label{rhoinp}
  \rho(T) = \frac{20}{19}\rho^*_{th}(T).
\end{equation}
 For this density we calculate the fraction of the pions in the condensate (further called condensate fraction). Of course, in the thermodynamic approximation this is $5\%$ by assumption.

Since the density of the condensate cannot exceed the total pion density, relation (\ref{conden}) implies that for $L\rightarrow 0$ the difference $E_0 - \mu$  tends to infinity and, consequently, the density of pions not in the condensate tends to zero\footnote{For models with discrete energy levels this follows from the fact that the energy difference between the first two energy levels $E_1-E_0 \rightarrow \infty$ when $L\rightarrow 0$.}. On the other hand, for $L\rightarrow \infty$ we must take $\mu\rightarrow E_0$ in order to get any condensate at all. Substituting $\mu = E_0$ into the formulae for the density of the non-condensate pions, and taking where necessary the limit $L \rightarrow \infty$, we get for each model the same result: the thermodynamic approximation. Thus, the remaining question is: how fast does the  condensate fraction decrease, with increasing $L$?

The density of pions not in the condensate is given for the free boundary conditions by formula (\ref{denter}) and for the other two models by the corresponding formula from statistical physics

\begin{equation}\label{rstarg}
  \rho ^*(T,\mu,L) = \frac{1}{L^3}\sum\frac{1}{e^{\beta(E-\mu)}-1},
\end{equation}
where the summation is over all the states with energies higher than $E_0$. From this point on, the densities are average densities, since we are working at fixed chemical potential and not at fixed number of particles, which makes a difference when the thermodynamic approximation does not hold. The total density of pions is

\begin{equation}\label{rhotot}
  \rho(T) = \rho_c(T,\mu,L) + \rho^*(T,\mu,L).
\end{equation}
This formula, whith a suitable redefinition of $\rho^*$ for the free boundary conditions, is valid for all our models, which implies that $\mu = \mu(T,L)$. Formula (\ref{conden}) yields the relation

\begin{equation}\label{}
  \mu(T,L) = E_0 - T\log\left(1 + \frac{1}{L^3\rho_c(T,\mu,L)}\right).
\end{equation}
Using equation (\ref{rhotot}) to elimiate $\rho_c$ and substituting the result into (\ref{rstarg}) one obtains an equation for $\rho^*(T,\mu(T,L),L)$, which can be easily solved by iterations, or by trial and error.

Let us define now the three models. In the model with periodic boundary conditions,

\begin{equation}\label{}
  E(n_1,n_2,n_3) = \sqrt{m^2 + \frac{4\pi^2}{L^2}s(n_1,n_2,n_3)},
\end{equation}
where

\begin{equation}\label{}
  s(n_1,n_2,n_3) = n_1^2 + n_2^2 + n_3^2
\end{equation}
and $n_1,n_2,n_3$ can be any integers. Thus, the ground state energy is $E_0 = m$.

In the model with wave functions vanishing at the boundaries of the cube,

\begin{equation}\label{}
  E(n_1,n_2,n_3) = \sqrt{m^2 + \frac{\pi^2}{L^2}s(n_1,n_2,n_3)}
\end{equation}
and $n_1,n_2,n_3$ are any positive integers. Thus, the ground state energy is

\begin{equation}\label{}
  E_0 =  \sqrt{m^2 + \frac{3\pi^2}{L^2}}.
\end{equation}

In the model with free boundary conditions

\begin{equation}\label{}
  E(\textbf{p}) = \sqrt{m^2+\textbf{p}^2},
\end{equation}
the integration as in (\ref{denter}) replaces the summation over $n_1,n_2,n_3$ and the ground state energy is $E_0 = m$.

The condensate fractions for the three models are compared with each other and with the thermodynamic approximation in Fig. 1. It is seen that, according to these models, the finite volume corrections are large and strongly model-dependent.

\begin{figure}[tpb]
\centering
\includegraphics{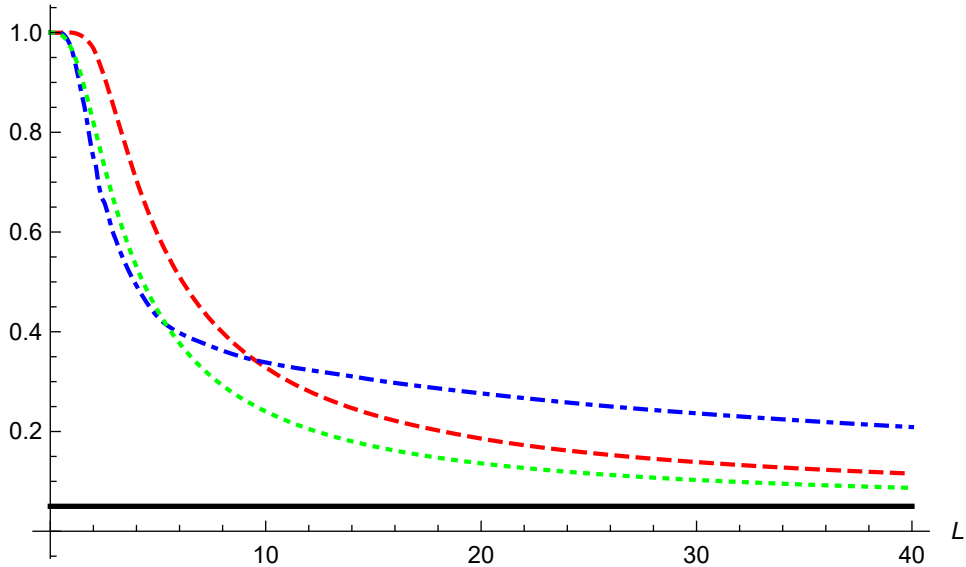}
\caption{Condensate fractions at $T = 100$MeV: continuous line - thermodynamic approximation (5\% by assumption), dashed line - periodic boundary conditions, dot-dashed line - wave functions vanishing at the boundaries, dotted line - free boundary conditions.}
\end{figure}

\section{Discussion and conclusions}

Phase transitions, with the accompanying discontinuities in the thermodynamic parameters and functions, can occur only in the thermodynamic limit. When the volume is finite and deviations from the thermodynamic limit are of interest, the phases can be defined only by analogy. It is natural (see \cite{BEF,BEG,AMV,BEG2}) to define the Bose-Einstein condensate as the set of particles in the single particle ground state.

Our calculations for the three models of pions condensing in a cubic box with edge length $L$ yield the following conclusions concerning the condensate fraction.
\begin{itemize}
\item With increasing $L$, the condensate fraction decreases from one at $L\rightarrow 0$   to the thermodynamic limit.
\item The decrease towards the thermodynamic limit is very slow. For a cubic box with a volume of  $L^3 = 6.4\cdot 10^4$fm$^3$, the corrections to the thermodynamic limit are still by factors of about  two and more. Thus, in practical calculations they must be taken into account.
\item The deviations from the thermodynamic limit depend strongly on the boundary conditions chosen. For example, for the ratios of the exactly calculated condensate fractions to the corresponding fraction in the thermodynamic approximation, at $L = 40$fm, we find: $2.3$ for the periodic boundary conditions, $4.2$ for the wave functions vanishing at the boundary and $1.7$ for the free boundary conditions. The corresponding numbers for $T=150$MeV are: $1.8$, $3.4$ and $1.4$. Thus, the corrections to the thermodynamic limit cannot be calculated reliably without specifying the choice of the boundary conditions.
\end{itemize}

\textbf{\emph{Acknowledgement}}
This work was partly supported by the Polish National Science Center (NCN) under grant DEC-2013/09/B/ST2/00497. The author thanks W. Florkowski for helpful comments.

\end{document}